\documentstyle[prd,aps,floats]{revtex}
\begin{document}
\draft
\preprint{BROWN-HET-1396}
\input epsf \renewcommand{\topfraction}{0.8}
\twocolumn[\hsize\textwidth\columnwidth\hsize\csname
@twocolumnfalse\endcsname

\title{Dynamical Relaxation of the Cosmological Constant and Matter
Creation in the Universe}

\author{Robert Brandenberger~$^{1,2,3}$ and Anupam Mazumdar~$^3$}

\address{$^1$~Department of Physics, Brown University, Providence, 
Rhode Island 02912, USA\\
$^2$~Perimeter Institute, 35 King St. North, Waterloo, ON, N2J 2W9, Canada,\\
$^3$~Department of Physics, McGill University, \\
3600 University Road, Montr\'eal, QC, H3A 2T8, Canada\\}
\maketitle
\begin{abstract}
In this Letter we discuss the issues of the graceful exit
from inflation and of matter creation in the
context of a recent scenario \cite{RHBrev} in which the back-reaction
of long wavelength cosmological perturbations induces a negative
contribution to the cosmological constant and leads to a dynamical
relaxation of the bare cosmological constant. The initially large
cosmological constant gives rise to primordial inflation, during
which cosmological perturbations are stretched beyond the Hubble
radius. The cumulative effect of the long wavelength fluctuations 
back-reacts on the background geometry in a form which corresponds
to the addition of a negative effective cosmological constant to the
energy-momentum tensor. In the absence of an effective 
scalar field driving inflation, whose decay can reheat the Universe,
the challenge is to find a mechanism which produces matter at the
end of the relaxation process. In this Letter, we point out that the decay
of a condensate representing the order parameter for a ``flat'' direction 
in the field theory moduli space can naturally provide a matter 
generation mechanism. The order parameter is displaced from its
vacuum value by thermal or quantum fluctuations, it is frozen until
the Hubble constant drops to a sufficiently low value, and then  
begins to oscillate about its ground state. During the period
of oscillation it can decay into Standard Model particles similar to 
how the inflaton decays in scalar-field-driven models of inflation. 
\end{abstract}

\pacs{BROWN-HET-1396}
\vskip2pc]

\section{Introduction}

The cosmological constant problem remains one of the most important 
problems of theoretical physics. The problem is to find an
explanation for the fact that the value of the cosmological constant 
in the present Universe - based on observational upper bounds on
the contribution of the cosmological constant to the ``energy''
content of the present Universe - is many orders of magnitude 
smaller than what is predicted according to theoretical estimates for
the vacuum energy (which acts like a cosmological constant). In 
non-supersymmetric theories, the mismatch is by a factor of
$10^{-120}$ (see e.g. \cite{Weinberg,Press} for reviews).  
Recent data, both from supernovae observations \cite{SNproject}
and also from cosmic microwave anisotropy
measurements (see \cite{WMAP} for the most recent observational
results) indicate that the Universe is right now entering a new
stage of acceleration, indicating the presence of something
that acts as an effective cosmological constant with the value 
$\Lambda_{eff} \sim 10^{-120} M_{p}^4$. Thus, there now appear
to be two aspects of the cosmological constant problem, firstly
why the cosmological constant is so tiny compared to theoretical
expectations (the ``old'' cosmological constant problem - using
Weinberg's language \cite{Weinberg2}) and the coincidence problem
of why the remnant cosmological constant is becoming visible
precisely at the present time of cosmic history (the ``new''
cosmological constant problem) and why it does not exactly vanish.

Independent studies of the back-reaction effects of long-wavelength
gravitational waves \cite{WT} and of long-wavelength cosmological
perturbations \cite{ABM} on the background geometry of space-time
have led to a scenario \cite{WT2,RHBrev} in which these fluctuations
lead to a dynamical relaxation of an initially large bare
cosmological constant. The large initial cosmological constant
leads to a period of primordial inflation. During this period,
quantum vacuum fluctuations of both gravitons and scalar metric
fluctuations are stretched beyond the Hubble radius, thus
generating a large phase space of long-wavelength modes (see 
e.g. \cite{MFB,RHBrev2} for reviews of the theory of cosmological
fluctuations). It can be shown that these modes have a back-reaction
effect on the local geometry which is analogous to that of a negative
cosmological constant (this effect is not physically measurable
in models with only one matter field \cite{Unruh,AW,GB1}, but it
is physically measurable in models with two or more matter fields
\cite{GB2}). This opens up the possibility that back-reaction
may lead to a dynamical relaxation of the bare cosmological constant.
What has been shown so far is only the perturbative onset of the
relaxation process (see e.g. \cite{WT3} for a modeling of
non-perturbative effects), but if the relaxation mechanism holds beyond
perturbation theory, it will lead \cite{RHBrev} to the inevitable
prediction of a dynamical scaling solution
\begin{equation} \label{fixed}
\Omega_{\Lambda_{eff}}(t) \sim 1 \, ,
\end{equation}
(where $\Omega_X = \rho_x / \rho_c$ is a measure of the contribution
of $X$ to the closure density $\rho_c$ of the Universe, $\rho_X$
denoting the effective energy density in some $X$ ``matter'')
for the effective cosmological constant, valid at all sufficiently
late times $t$. Thus, this dynamical relaxation mechanism would 
automatically address both the old and the new cosmological constant
problems.

The reason why we obtain the dynamical fixed point 
solution (\ref{fixed}) is as follows: as the phase
of inflation proceeds, the phase space of infrared modes 
builds up. Since long-wavelength fluctuations are frozen,
the contribution of an individual mode to the effective
cosmological constant does not decrease in absolute magnitude.
Thus, the total back-reaction effect builds up gradually,
and the effective cosmological constant, which is the sum
of the positive bare cosmological constant and the induced
negative back-reaction term, grows. However, the sum cannot
drop to zero, since before this happens the energy density
$\rho_{\Lambda}$ corresponding to the {\it effective} cosmological
constant will drop below the matter energy density. As soon
as this happens, inflation will end, the phase space of infrared
modes will cease to increase, and the back-reaction contribution
to $|\Lambda_{eff}|$ levels off. Since the Universe is still
expanding, the matter energy density $\rho_{m}$ (where ``matter''
here stands for both cold matter and radiation) continues to
decrease, thus allowing $\rho_{\Lambda}$ to start dominating
again, enabling the back-reaction effect to again increase
in strength. Thus, as explained in detail in \cite{RHBrev}, we
expect that at all sufficiently late times $\rho_{\Lambda} / \rho_m$
will be oscillating in time about the value $0.5$. This is
a ``tracking'' behavior very similar to what is postulated
in tracking quintessence models \cite{quint} \footnote{Note
that a similar scenario where the effective cosmological constant
fluctuates and whose associated energy density always tracks
that of matter emerges from the causal set approach to quantum
gravity \cite{Sorkin,SorDod}. Whereas in our approach the
effective cosmological constant must remain positive, in the
causal set approach its sign fluctuates.}.

However, since during the period of inflation the energy
density in cold matter and radiation decreases exponentially,
there needs to be a mechanism to provide a ``graceful exit'' from
inflation and to reheat the Universe. Otherwise,
the matter energy density during the late time ``tracking'' period
will be much too low (note that in order for back-reaction of
infrared modes to have an important effect, the period of inflation
has to be very long - see \cite{ABM} for the precise numbers).  

In this Letter we propose a solution to this problem which
is completely natural from the point of view of the
Minimal Supersymmetric Standard Model (MSSM) and many other models
of particle physics beyond the Standard Model. These models
contain light scalar field (massless in the absence of supersymmetry
breaking). During the period of inflation, these fields will
form condensates which are displaced from the absolute minima
of the respective potentials (after supersymmetry breaking). The
displacements may be due to the pre-inflationary initial conditions,
or they may develop during inflation via quantum fluctuations. 
During inflation the
dynamical evolution of a condensate is effectively frozen, but after
inflation its dynamics will become important. Specifically, at
a value of the Hubble expansion rate $H$ determined by the
potential of the condensate (see Section 3 for details) - much
smaller than the initial Hubble expansion rate set by the value
of the bare cosmological constant - 
the energy density of the condensate starts to dominate over the
energy density associated with the effective cosmological constant, 
and the inflationary period will end. At a comparable time,
the condensate will become ``unfrozen'' and begin to oscillate about
its ground state. The decay of the condensate into radiation will
reheat the Universe by a mechanism very similar to what happens
in the final decay of the inflaton condensate in scalar-field-driven
inflation. 

Particle physics provides ready examples of how to implement
our ideas. For example, in the MSSM, $F$-and $D$-flat
directions can form a condensate during inflation (for cosmological
consequences of condensates, see~\cite{Enqvist}). 
Since the condensate can
carry MSSM charges, its decay will automatically give rise
to a relativistic bath of Standard Model degrees of freedom. 
This decay should reheat the Universe to temperature higher
than that when the synthesis of light nuclei become
important~\cite{Enqvist:2002rf}. Note that, in this context,
our mechanism provides a natural explanation for the reheating of 
the Universe into a thermal bath with the
right degrees of freedom required for big bang nucleosynthesis (BBN).

We mention two more ways to realize our scenario. The
first is by making use of
light Majorana neutrinos. In this case, the supersymmetric partner of
right handed Majorana neutrino, the sneutrino, can act as a
condensate, provided the lightest of the sneutrinos is lighter than
the Hubble expansion rate during inflation~\cite{Mazumdar,Lorenzana}.
The second utilizes the moduli fields (scalar fields which are
massless before supersymmetry breaking) which arise in the
low energy limit of string theories (see e.g. \cite{moduli}) in which
the extra spatial dimensions are compactified.

The outline of this paper is as follows. We first review the
mechanism by which cosmological fluctuations back-react on the
expansion rate of the Universe. In Section 3 we discuss the
dynamics of the condensate and its decay. In Section 4 we
conclude with a discussion of some of the challenges for late time
cosmology in our scenario.


\section{Review of the Back-Reaction Formalism}

It has been know for a long time \cite{Brill}
that gravitational waves carry energy and momentum and can have
a back-reaction effect on the background in which they propagate.
In a similar way, we expect cosmological fluctuations (scalar
metric fluctuations) to be described by an energy-momentum
tensor which influences the evolution of the background.

The formalism to discuss this back-reaction in the case
of cosmological perturbations was initially developed in
\cite{ABM}. The idea is to consider small fluctuations of the
metric and the matter fields about a homogeneous and isotropic background,
to insert these fluctuations into the Einstein equations and expand
these equations to second order in the small parameter 
$\epsilon$ which parameterizes
the amplitude of the fluctuations. Due to their non-linearity, the
Einstein equations are not satisfied at second
order, and it is necessary to add back-reaction terms which are
quadratic in $\epsilon$. In particular, one needs to add a
second order correction to the zero mode of the metric. The
sum of the background metric plus the quadratic zero mode
correction defines a new homogeneous metric $g_{\mu\nu}^{(0,br)}$
which takes into account the effects of back-reaction to this
order. This new metric obeys the modified equations
\begin{equation}
G_{\mu\nu}(g_{\alpha\beta}^{(0,br)}) = 8 \pi G \left[T_{\mu\nu}^{(0)}
+\left\langle T^{(2)}_{\mu\nu} - {1 \over {8 \pi G}}G^{(2)}_{\mu\nu}\right
\rangle\right]\,,
\end{equation}
where $G_{\mu \nu}$ indicates the Einstein tensor, $T_{\mu \nu}$ is the
energy-momentum tensor of matter, the pointed brackets stand for spatial 
averaging, and the superscripts indicate the order in perturbation theory.
The terms inside the pointed brackets form what is called the
effective energy-momentum tensor $\tau_{\mu \nu}$ of cosmological
perturbations \footnote{Note that even if inflation is driven
by the bare cosmological constant, we need to introduce a matter
field in order to have scalar metric fluctuations. For simplicity,
we shall introduce a scalar field $\varphi$ which is slowly rolling
during the period of inflation (but which has a negligible effect
on the background dynamics, at least initially).} Note that all
Fourier modes of the gravitational potential $\Phi$ contribute to
the effective energy-momentum tensor for back-reaction: the
effect of fluctuations is cumulative. Thus, back-reaction can
be important even if the relative magnitude of each metric
fluctuation mode is small (as observations indicate) as long as
there is a sufficiently large phase space of infrared modes, i.e.
as long as the period of primordial inflation is sufficiently long.

For long wavelength fluctuations, the terms involving the metric
fluctuations are dominant. It was shown in \cite{ABM} that the
back-reaction equation is covariant under linear space-time
coordinate transformations. Thus, to simplify the analysis,
we can work in a convenient gauge, namely
Longitudinal gauge in which the metric is diagonal
\begin{equation}
ds^2=(1+2\Phi)dt^2-a(t)^2(1-2\Phi)\delta_{ij}dx^{i}dx^{j}\,,
\end{equation}
where $\Phi(x,t)$ is the metric perturbation and $a(t)$ is the scale factor.

In addition to the metric fluctuations, there are matter fluctuations.
One matter fluctuation degree of freedom (the ``adiabatic'' matter
perturbation) is related to the metric
potential $\Phi$ via the Einstein constraint equation. For example, in
the case of a matter model with a single scalar field $\varphi$, 
the matter field fluctuation $\delta \varphi$ is related to $\Phi$ via
\begin{equation}
\dot \Phi +H\Phi = 4\pi G\dot \varphi _0\,\delta \varphi \,.  \label{constr}
\end{equation}
Making use of the background equation of motion for the inflaton
field $\varphi$ during the slow rolling phase, one can replace
${\dot {\varphi}}$ in (\ref{constr}) by the scalar field potential
$V(\varphi)$ and its derivative. Since on large scales 
$\dot \Phi \simeq 0$, the fluctuation variables $\delta \varphi$
and $\Phi$ are directly related, and all terms in the effective
energy-momentum tensor $\tau_{\mu \nu}$ can be expressed as
functions of the background variables multiplying $<\Phi ^2>$.

In hindsight, it is easy to understand why the back-reaction of
infrared modes of cosmological perturbations acts as a negative
cosmological constant. For long wavelength fluctuations, all terms
in the effective energy-momentum tensor involving spatial derivatives
are negligible. Since on long wavelengths the amplitude of the
metric fluctuation variable $\Phi$ is frozen (see e.g. \cite{MFB}),
all terms involving temporal derivatives of the fluctuation variables
also vanish. The only terms which survive are gravitational potential
energy terms. Since matter fluctuations produce negative potential
wells, and since for infrared modes the negative gravitational energy
dominates over the positive matter energy, the total effective energy
density is negative, and the associated energy-momentum tensor takes
has the equation of state of a potential energy term, i.e. that of
a cosmological constant. 

In the long-wavelength limit for fluctuations, and the slow-roll
approximation for the dynamics of the matter field $\varphi$ (with
potential energy function $V(\varphi)$),
the expression for the effective energy density $\rho_{eff}$
becomes \cite{ABM}
\begin{equation} 
\rho _{br}\equiv \tau _0^0\cong \left( 2\,{\frac{{V^{\prime \prime }V^2}}{{%
V^{\prime }{}^2}}}-4V\right) <\Phi ^2>\,,  \label{tzerolong}
\end{equation}
which is typically negative for shallow potentials.
The effective energy-momentum tensor $\tau_{\mu \nu}$ takes the 
form of a {\it negative cosmological constant}
\begin{equation}
\label{result}
p_{br}=-\rho _{br} \,\,\, {\rm with} \,\,\, \rho_{br} < 0 \, .
\end{equation}
A crucial observation is that the magnitude of $\rho_{br}$ 
increases as a function of time. This is because, in an inflationary 
Universe, as time increases more and more wavelengths become longer than the 
Hubble radius and begin to contribute to $\rho_{br}$. 

However, as first pointed out in \cite{Unruh} (see also \cite{Niayesh})
and later shown
rigorously in \cite{AW} and \cite{GB1}, in single matter field
models the back-reaction of the dominant infrared terms is not
physically measurable. It can be locally masked by a time translation.
This can be seen by studying the back-reaction of long wavelength
cosmological perturbations on a local measure of the expansion rate,
and expressing the result in terms of a local clock. However, if
there is an additional matter field present which can be used to
define a physical clock (this field may be viewed as describing the
temperature of the cosmic microwave background), then the leading
back-reaction terms described here are real, as recently shown
in \cite{GB2} (see also \cite{AW2} for another model which demonstrates
that infrared back-reaction is ``for real''). 
Since our model involves at least one additional
matter field (namely the condensate), the infrared back-reaction
will be physically measurable.

Since it is the total metric and not the background metric only
which determines observables, it was suggested \cite{RHBrev} 
on the basis of the above results that the effective 
cosmological constant at time $t$ is given by
\begin{equation}
\label{result1}
\Lambda_{eff}(t)=\Lambda_{0}+\rho_{br}\, ,
\end{equation}
where the second term on the right hand side is negative and has
an absolute value which is increasing in time. Since mode by
mode the magnitude of the square of the fluctuation amplitude
which enters the expectation value $<\Phi ^2>$ is tiny, 
the back-reaction contribution to the effective cosmological 
constant is very small compared to 
the bare cosmological constant $\Lambda_{0}$ during the early
stages of inflation. However, if inflation lasts long enough (as
will be the case in many inflationary models in the class of
``chaotic'' inflation \cite{chaotic}), then, before the
homogeneous scalar field $\varphi$ has ended the slow-rolling phase, the
back-reaction contribution to the effective cosmological constant 
will cancel the initial bare value $\Lambda_{0}$. 
However, as described in the Introduction, the energy
density associated with $\Lambda_{eff}(t)$ can in fact never
become negative, and at late times one will obtain a dynamical
scaling solution in which the energy density associate with
$\Lambda_{eff}(t)$ tracks the matter energy density.


\section{Condensate Dynamics and Reheating}

As mentioned in the Introduction, in order that the back-reaction
scenario for dynamically relaxing the bare cosmological constant
produce a Universe compatible with the one we observe today,
there needs to be some component of matter whose energy
density does not red-shift significantly during primordial
inflation. Condensates which are frozen during
inflation can naturally play this role. As mentioned in the
Introduction, the existence of such condensate (or moduli) fields
is a generic feature of many particle physics theories beyond
the Standard Model.

In this paper we assume a generic flat direction field,
$\chi$, which is lifted by a small mass $m_{\chi}$. We assume that the
mass is small compared to the Hubble expansion during the early
stages of inflation.
Therefore $\chi$ is free to fluctuate along this flat direction.
Because inflation smoothes out all gradients, only the homogeneous
condensate mode survives \footnote{Fluctuations are continuously
generated in the ultraviolet, but if the Hubble constant is
decreasing, the spectrum of fluctuations will be blue, and the
amplitude of the short wavelength fluctuations will rapidly 
become tiny compared to the size of the ``zero mode''. Thus,
we will focus on the mode of $\chi$ which is quasi-homogeneous.} 
The fluctuations of the condensate do play a role, however, in
that they produce the entropy fluctuations which are crucial
\cite{FB,GB2} in order that the back-reaction effect be
physically measurable.

While the energy density of the condensate is much smaller than the
energy in the bare cosmological constant, the Hubble expansion
rate $H(t)$ is not influenced by $\chi$. If $H(t)$ were
constant, then the condensate $\chi$ would be slowly rolling
provided $m_{\chi} \ll H$, i.e. the acceleration term
in the equation of motion for $\chi$
\begin{equation}
\ddot \chi + 3H(t)\dot\chi + m_{\chi}^2\chi = 0 
\end{equation}
would be negligible. However, in our case, as the back-reaction
effect grows, the time-dependence of $H(t)$ will become important
and the slow-rolling approximation has to be modified even while
$m_{\chi} \ll H$. We make the {\it modified slow rolling ansatz}
\begin{equation} \label{SRansatz}
{\dot \chi} = - {{\alpha m_{\chi}^2} \over {3 H(t)}} \chi \, ,
\end{equation}
where $\alpha$ is a constant to be determined ($\alpha = 1$ if
$H(t)$ is independent of $t$). With this ansatz,
the second derivative of $\chi$ now contains two terms
\begin{equation}
{\ddot \chi} = \left({{m_{\chi}^2} \over {3 H(t)}}\right)^2 \chi
+ {{m_{\chi}^2} \over {3 H(t)^2}} {\dot H} \chi \, .
\end{equation}
While $m_{\chi} \ll H$, the first term is negligible. Using the
background Friedmann-Robertson-Walker equations to substitute
for ${\dot H}$, we then find
\begin{equation}
\alpha = {1 \over {1 + {{1 + w} \over 2}}} \, ,
\end{equation}
where $w(t)$ labels the effective equation of state of the background
(in homogeneous isotropic cosmology, $w = p / \rho$, where $p$ is
the pressure and $\rho$ is the energy density).
 
Integrating Equation (\ref{SRansatz}) from some given time $t_0$ to
time $t$ then yields the following solution for $\chi(t)$:
\begin{equation}
\chi(t) = \chi(t_0) {\rm exp}\left(- \alpha m_\chi^2 \int_{t_0}^t H^{-1} dt
\right) \, .
\end{equation}
For a fixed value of $w$, the time integral in the exponent for
$t \gg t_0$ yields the value $\beta / 2 H(t)^2$, where
\begin{equation}
\beta = {2 \over {3(1 + w)}} \, ,
\end{equation}
from which it
follows that $\chi(t)$ is effectively frozen until $H(t) = m_{\chi}$.

Once $H(t)$ falls below $m_{\chi}$, the condensate $\chi$ will
start oscillating about $\chi = 0$ in a manner analogous to how
the inflaton starts oscillating about the minimum of its potential
at the end of inflation. If the initial value of the condensate
is smaller than the Planck mass
$m_{pl}$, the energy density during the initial
period of oscillation of $\chi$ will still be dominated by the
effective cosmological constant (since $\rho_{\chi} < \rho_{\Lambda}$).
However, given initial conditions for $\chi$ as in the chaotic inflation 
scenario \cite{chaotic}, one would expect $\chi(t_0) \geq m_{pl}$
and thus the energy density in $\chi$ would begin to dominate before
or when $\chi$ starts oscillating, and
the Hubble expansion rate would be determined by the condensate 
dynamics, which mimics a time-averaged equation of state of dust.
 
Let us imagine (in analogy with early studies of reheating in
inflation \cite{DL,AFW}) that the condensate decays perturbatively into
light fermions with a decay width $\Gamma\sim g^2 m_{\chi}$. Then,
the energy density in $\chi$ will be transferred effectively to
matter only when 
\begin{equation}
\Gamma \sim H \, . 
\end{equation}
The reheating temperature $T_{reh}$
(the temperature of ordinary matter after the energy transfer is
completed) can then be estimated via
\begin{equation}
T_{reh}^4 \sim {3 \over {8 \pi G}} H_{dec}^2 \, ,
\end{equation}
where $H_{dec}$ is the value of $H$ when the energy transfer takes
place. This yields
\begin{equation}
T_{reh} \sim \left({3\over {8 \pi}} \right)^{1/4} g m_{pl}^{1/2} 
m_{\chi}^{1/2} \, ,
\end{equation}
where $m_{pl}$ is the Planck mass.
Note that for sufficiently small values of $m_{\chi} \ll 10^7$~GeV 
and for $g \leq 10^{-4}$, it
is possible to obtain a reheat temperature below $10^{9}$~GeV
(similar constraints were obtained in \cite{Lorenzana,Kasuya}) which
would render our scenario safe from the potential problem of 
overproduction of thermal and non-thermal gravitinos \cite{Ellis}. 

On the other hand, for a condensate whose potential is dominated
by the mass term (as we assume here), it is likely that the
initial stages of the energy transfer from the condensate to
matter will proceed via the parametric resonance
instability \cite{TB}. This occurs both if $\chi$ is coupled to
bosons \cite{KLS} or to fermions \cite{GK}.
In this case, the energy density of matter after energy
transfer corresponds to a matter temperature $T_{max}$ of
\begin{equation} 
T_{max} \sim m_{\chi}^{1/2} m_{pl}^{1/2} \, .
\end{equation}
In this case, the upper bound on $m_{\chi}$ to be safe from
the gravitino over-abundance problem is lower than if the
decay occurs perturbatively.

In both cases, the decay of the condensate 
naturally provides us a hot thermal Universe required for BBN. If the
condensate carries some local or global quantum number, such as
baryonic charge, then it is possible that the decay yields
a baryon to entropy ratio of $10^{-10}$ which is required for 
successful BBN (see e.g. \cite{Mazumdar,Lorenzana}).

\section{Conclusion}
 
In this Letter we discussed a natural way of gracefully exiting
the period of primordial inflation and of reheating the
Universe in the scenario in which the back-reaction of
long-wavelength cosmological fluctuations dynamically
relaxes a bare cosmological constant which provides a
period of primordial inflation. The mechanism makes
use of the dynamics of a condensate field $\chi$ which has 
a small mass $m_{\chi}$ and which has a quasi-homogeneous mode
which is excited during the initial stage of inflation. This
condensate is frozen until the effective Hubble parameter
drops below the condensate mass. 

The main role of the condensate is to provide a form of matter
energy which is not red-shifted exponentially during the period
of primordial inflation. When $H(t)$ drops below $m_{\chi}$,
the energy density of the Universe becomes dominated by the
condensate field, thus terminating the period of primordial
inflation naturally. At that time, the condensate $\chi$ begins to oscillate
about $\chi = 0$, and its decay will reheat the Universe
to a temperature determined by $m_{\chi}$ and by dimensionless
coupling constants which describe the coupling of $\chi$ to
regular matter.

Candidates for the condensate are moduli fields of string
compactifications, MSSM flat directions or sneutrino
condensates. Such condensates are naturally excited in
the context of inflationary cosmology. An added bonus of
our scenario is that the decay of the condensate, which
starts once the condensate unfreezes, automatically provides
a hot thermal bath of Standard Model particles, a prerequisite
for successful big bang nucleosynthesis. We
show that it is possible to reheat the Universe to below $10^{9}$~GeV if
the Yukawa coupling has a value $g \sim 10^{-4}$ and if the 
effective mass of the condensate satisfies
$\sim 10^7$~GeV. For smaller masses, the reheat temperature
can be further lowered, and therefore there is no danger of over-producing
thermal/non-thermal gravitinos.

Our work is one step forwards towards connecting the scenario
of \cite{RHBrev} with a realistic late-time cosmology. However,
a lot more work needs to be done in this area. In particular,
it is important to study the effects of an effective cosmological
constant which is oscillating about $\Omega_{\Lambda} = 1/2$ (the
dynamical fixed point solution described in Section 2) on
big bang nucleosynthesis, structure formation, and the evolution
of cosmic microwave background anisotropies \footnote{Any model
with an oscillating effective cosmological constant, such as the
model of \cite{Sorkin}, will have to address these issues.}.
Work on these problems is in progress \cite{Anupam}.

\subsection*{Acknowledgments}

R.B. is supported in part by the US Department of Energy under Contract
DE-FG02-91ER40688, TASK~A. He thanks the Fields Institute for
Mathematical Sciences and the Perimeter Institute for hospitality and
support during the final stages of this project.
A.M. is a CITA National Fellow. His work is also supported in
part by NSERC (Canada) and by the Fonds de Recherche sur la
Nature et les Technologies du Qu\'ebec.

\vskip20pt


\end{document}